\documentclass[showpacs,10pt,twocolumn,prx]{revtex4-1}
\usepackage{amsmath}
\usepackage{amssymb}
\usepackage{graphics}
\usepackage{epsfig}
\usepackage{color}
\usepackage{lineno}

\begin{document}


\title{Superconductivity and normal-state properties of kagome metal RbV$_{3}$Sb$_{5}$ single crystals}
\author{Qiangwei Yin$^{\dag}$, Zhijun Tu$^{\dag}$, Chunsheng Gong$^{\dag}$, Yang Fu, Shaohua Yan, and Hechang Lei}
\email{hlei@ruc.edu.cn}
\affiliation{Department of Physics and Beijing Key Laboratory of Opto-electronic Functional Materials $\&$ Micro-nano Devices, Renmin University of China, Beijing 100872, China}

\date{\today}

\begin{abstract}
We report the discovery of superconductivity and detailed normal-state physical properties of RbV$_{3}$Sb$_{5}$ single crystals with V kagome lattice. RbV$_{3}$Sb$_{5}$ single crystals show a superconducting transition at $T_{c}\sim$ 0.92 K. Meanwhile, resistivity, magnetization and heat capacity measurements indicate that it exhibits anomalies of properties at $T^{*}\sim$ 102 - 103 K, possibly related to the formation of charge ordering state. When $T$ is lower than $T^{*}$, the Hall coefficient $R_{\rm H}$ undergoes a drastic change and sign reversal from negative to positive, which can be partially explained by the enhanced mobility of hole-type carriers. In addition, the results of quantum oscillations show that there are some very small Fermi surfaces with low effective mass, consistent with the existence of multiple highly dispersive Dirac band near the Fermi energy level.
\end{abstract}


\maketitle

Two-dimensional (2D) kagome lattice composed of corner-sharing triangles and hexagons is one of the most studied systems in the last decades due to its unique structural feature. On the one hand, if only the spin degree of freedom is considered, insulating magnetic kagome materials can host some exotic magnetism ground states, like quantum spin liquid state, because of the nature of strongly geometrical frustration for kagome lattice \cite{Balents,Shores,HanTH,FuM}. 
On the other hand, when the charge degree of freedom becomes dominant (partial filling), the band topology starts to manifest its features in kagome metals, such as nontrivial Dirac points and flat band in the band structure \cite{KangM,LiuZ,KangM2}. More interestingly, when both of spin and charge degrees of freedom exist, many of exotic phenomena appear in the correlated magnetic kagome metals. For example, in ferromagnetic Fe$_{3}$Sn$_{2}$ and TbMn$_{6}$Sn$_{6}$ kagome metals, due to the spin orbital coupling and broken of time reversal symmetry, the Chern gap can be opened at the Dirac point, leading to large anomalous Hall effect (AHE), topological edge state and large magnetic-field tunability \cite{YeL,YinJX,YinJX2}. Moreover, antiferromagnetic Mn$_{3}$Sn and ferromagnetic Co$_{3}$Sn$_{2}$S$_{2}$ kagome metals exhibit large intrinsic AHE, which is related to the existence of Weyl node points in these materials \cite{Kuroda,LiuE,WangQ}. 

Besides the intensively studied magnetic kagome metals, other correlation effects and ordering states in partially filled kagome lattice have also induced great interests. Theoretical studies suggest that the doped kagome lattice could lead to unconventional superconductivity \cite{Balents,Anderson,Ko,WangWS,Kiesel}. Especially, when the kagome lattice is filled near van Hove filling, the Fermi surface (FS) is perfectly nested and has saddle points on the $M$ point of Brillouin zone (BZ) \cite{WangWS}. Depends on the variations of on-site Hubbard interaction $U$ and Coulomb interaction on nearest-neighbor bonds $V$, the system can develop different ground states, including unconventional superconductivity, ferromagnetism, charge bond order and charge density wave (CDW) order and so on \cite{WangWS,Kiesel}. However, the realizations of superconducting and charge ordering states are still scarce in kagome metals.

In very recent, a novel family of kagome metals AV$_{3}$Sb$_{5}$ (A = K, Rb and Cs) was discovered \cite{Ortiz1}. Among them, KV$_{3}$Sb$_{5}$ and CsV$_{3}$Sb$_{5}$ exhibit superconductivity with transition temperature $T_{c}=$ 0.93 and 2.5 K, respectively \cite{Ortiz2,Ortiz3}. The proximity-induced spin-triplet superconductivity was also observed in Nb/KV$_{3}$Sb$_{5}$ devices \cite{WangY}. More importantly, theoretical calculations and angle-resolved photoemission spectroscopy (ARPES) demonstrate that there are several Dirac nodal points near the Fermi energy level ($E_{\rm F}$) with a non-zero $Z_{2}$ topological invariant in KV$_{3}$Sb$_{5}$ and CsV$_{3}$Sb$_{5}$ \cite{Ortiz1,Ortiz2,Ortiz3,YangSY}. Moreover, AV$_{3}$Sb$_{5}$ exhibits transport and magnetic anomalies at $T^{*}\sim$ 80 K - 110 K \cite{Ortiz1,Ortiz2,Ortiz3}. The X-ray diffraction (XRD) and scanning tunnelling microscopy (STM) measurements on KV$_{3}$Sb$_{5}$ and CsV$_{3}$Sb$_{5}$ indicate that there is a 2$\times$2 superlattice emerging below $T^{*}$, i.e., the formation of charge order (CDW-like state) \cite{Ortiz2,JiangYX}. Furthermore, the STM spectra show that this charge order has a chiral anisotropy, which can be tuned by magnetic field and may lead to the anomalous Hall effect (AHE) at low temperature even KV$_{3}$Sb$_{5}$ does not exhibit magnetic order or local moments \cite{YangSY,JiangYX,Kenney}.

Motivated by these studies, in this work, we carried out a comprehensive study on physical properties of RbV$_{3}$Sb$_{5}$ single crystals. We find that RbV$_{3}$Sb$_{5}$ shows a superconducting transition at $T_{c}\sim$ 0.92 K, which coexist with the anomalies of properties at $T^{*}\sim$ 102 - 103 K. This could be related to the emergence of charge ordering state. Below $T^{*}$, the transport properties change significantly, possibly rooting in the dramatic changes of electronic structure due to the formation of charge order. Furthermore, the analysis of low-temperature quantum oscillations indicates that there are small Fermi surfaces (FSs) with low effective mass in RbV$_{3}$Sb$_{5}$, revealing the existence of highly dispersive bands near the Fermi energy level $E_{\rm F}$.

Single crystals of RbV$_{3}$Sb$_{5}$ were grown from Rb ingot (purity 99.75\%), V powder (purity 99.9\%) and Sb grains (purity 99.999\%) using the self-flux method \cite{Ortiz2}. The eutectic mixture of RbSb and Rb$_{3}$Sb$_{7}$ is mixed with VSb$_{2}$ to form a composition with 50 at.\% Rb$_{x}$Sb$_{y}$ and 50 at.\% VSb$_{2}$ approximately. The mixture was put into an alumina crucible and sealed in a quartz ampoule under partial argon atmosphere. The sealed quartz ampoule was heated to 1273 K for 12 h and soaked there for 24 h. Then it was cooled down to 1173 K at 50 K/h and further to 923 K at a slowly rate. Finally, the ampoule was taken out from the furnace and decanted with a centrifuge to separate RbV$_{3}$Sb$_{5}$ single crystals from the flux. Except sealing and heat treatment procedures, all of other preparation processes were carried out in an argon-filled glove box in order to prevent the reaction of Rb with air and water. The obtained crystals have a typical size of 2 $\times$ 2 $\times$ 0.02 mm$^{3}$. RbV$_{3}$Sb$_{5}$ single crystals are stable in the air. 
XRD pattern was collected using a Bruker D8 X-ray diffractometer with Cu $K_{\alpha}$ radiation ($\lambda=$ 0.15418 nm) at room temperature. The elemental analysis was performed using the energy-dispersive X-ray spectroscopy (EDX). Electrical transport and heat capacity measurements were carried out in a Quantum Design physical property measurement system (PPMS-14T). The longitudinal and Hall electrical resistivity were measured using a five-probe method and the current flows in the $ab$ plane of the crystal. The Hall resistivity was obtained from the difference in the transverse resistivity measured at the positive and negative fields in order to remove the longitudinal resistivity contribution due to the voltage probe misalignment, i.e., $\rho_{yx}(\mu_{0}H)=[\rho_{yx}(+\mu_{0}H)-\rho_{yx}(-\mu_{0}H)]/2$. The $c$-axial resistivity was measured by attaching current and voltage wires on the opposite sides of the plate-like crystal. Magnetization measurements were performed in a Quantum Design magnetic property measurement system (MPMS3).


\begin{figure}
\centerline{\includegraphics[scale=0.28]{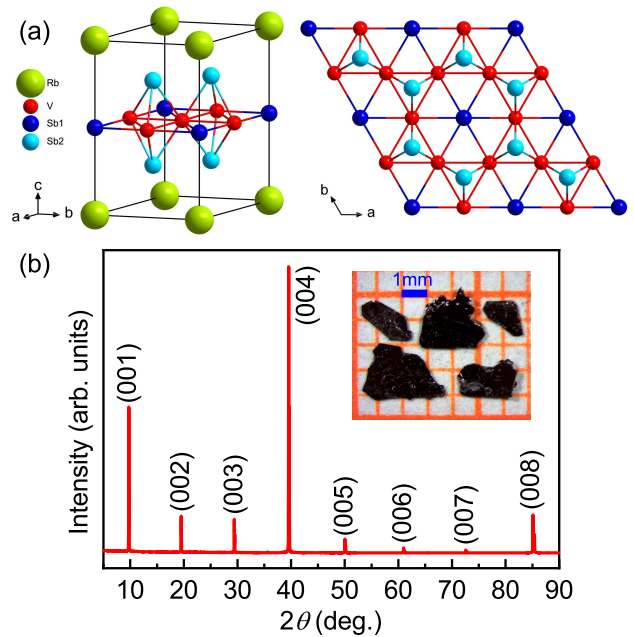}} \vspace*{-0.3cm}
\caption{(a) Crystal structure of RbV$_{3}$Sb$_{5}$. The big green, small red, medium blue and cyan balls represent Rb, V, Sb1 and Sb2 sites, respectively. (b) XRD pattern of a RbV$_{3}$Sb$_{5}$ single crystal. Inset: photo of typical RbV$_{3}$Sb$_{5}$ single crystals on a 1 mm grid paper.}
\end{figure}

As shown in the left panel of Fig. 1(a), RbV$_{3}$Sb$_{5}$ has a layered structure with hexagonal symmetry (space group $P6/mmm$, No. 191). It consists of Rb layer and V-Sb slab stacking along $c$ axis alternatively, isostructural to KV$_{3}$Sb$_{5}$ and CsV$_{3}$Sb$_{5}$ \cite{Ortiz1}. 
The key structural ingredient of this material is two-dimensional (2D) kagome layer formed by the V atoms in the V-Sb slab (right panel of Fig. 1(a)). There are two kinds of Sb sites and the Sb atoms at Sb1 site occupy at the centers of V hexagons when another Sb atoms at Sb2 site locate below and above the centers of V triangles, forming graphene-like hexagon layers.
The XRD pattern of a RbV$_{3}$Sb$_{5}$ single crystal (Fig. 1(b)) reveals that the crystal surface is parallel to the $(00l)$-plane. The estimated $c$-axial lattice constant is about 9.114 \AA, close to previously reported values \cite{Ortiz1}. The thin-plate-like crystals (inset of Fig. 1(b)) are also consistent with the layered structure of RbV$_{3}$Sb$_{5}$. The measurement of EDX by examination of multiple points on the crystals gives the atomic ratio of Rb : V : Sb = 0.90(6) : 3 : 5.07(4) when setting the content of V as 3. The composition of Rb is slightly less than 1, indicating that there may be small amount of Rb deficiencies in the present RbV$_{3}$Sb$_{5}$ crystals.

Fig. 2(a) exhibits the temperature dependence of in-plane resistivity $\rho_{ab}(T)$ and $c$-axial resistivity $\rho_{c}(T)$ of RbV$_{3}$Sb$_{5}$ single crystal from 2 K to 300 K. The zero-field $\rho_{ab}(T)$ exhibits a metallic behavior in the measured temperature range and the residual resistivity ratio (RRR), defined as $\rho_{ab}$(300 K)/$\rho_{ab}$(2 K), is about 44, indicating the high quality of crystals.
At $T^{*}\sim$ 103 K, the $\rho_{ab}(T)$ shows an inflection point and it is related to the onset of charge ordering transition \cite{Ortiz2,JiangYX}. It should be noted that the $T^{*}$ is higher than those in KV$_{3}$Sb$_{5}$ and CsV$_{3}$Sb$_{5}$ \cite{Ortiz2,Ortiz3}, implying that the relationship between $T^{*}$ and the lattice parameters (or ionic radius of alkali metal) is not monotonic. At $\mu_{0}H =$ 14 T, $\rho_{ab}(T)$ is insensitive to magnetic field when $T>T^{*}$ but the significant magnetoresistance (MR) appears gradually below $T^{*}$. On the other hand, the $\rho_{c}(T)$ has a much larger absolute value than the $\rho_{ab}(T)$. The ratio of $\rho_{c}/\rho_{ab}$ is about 7 at 300 K and increases to about 33 when temperature is down to 2 K, manifesting a significant 2D nature of RbV$_{3}$Sb$_{5}$. However this anisotropy is smaller than that in CsV$_{3}$Sb$_{5}$, which could be partially ascribed to the smaller interlayer spacing between two V-Sb slabs \cite{Ortiz2}. More importantly, in contrast to $\rho_{ab}(T)$, the $\rho_{c}(T)$ shows a remarkable upturn starting from $T^{*}$ with a maximum at about 97 K and this behavior is distinctly different from that in CsV$_{3}$Sb$_{5} $\cite{Ortiz2}. It suggests that 
the $\textbf{q}_{\rm CDW}$ in RbV$_{3}$Sb$_{5}$ might have a $c$-axial component, leading to the significantly gapped FS along the $k_{z}$ direction. Similar behavior has also been observed in PdTeI with CDW vector $\textbf{q}_{\rm CDW}$ = (0, 0, 0.396) \cite{LeiHC} and GdSi with spin density wave (SDW) vector $\textbf{q}_{\rm SDW}$ = (0, 0.483, 0.092) \cite{FengY}.
Fig. 2(b) exhibits the $\rho_{ab}(T)$ as a function of temperature below 1.3 K. It can be seen that there is a sharp resistivity drop appearing in the $\rho_{ab}(T)$ curve at zero field and it corresponds to the superconducting transition. The onset superconducting transition temperature $T_{c,\rm onset}$ determined from the cross point of the two lines extrapolated from the high-temperature normal state and the low-temperature superconducting state is 0.92 K with the transition width $\Delta T_{c}=$ 0.17 K. This $T_{c}$ is lower than that of CsV$_{3}$Sb$_{5}$ ($T_{c}\sim$ 2.5 K) but very close to that of KV$_{3}$Sb$_{5}$ ($T_{c}\sim$ 0.93 K) \cite{Ortiz2,Ortiz3}. 

\begin{figure}
\centerline{\includegraphics[scale=0.15]{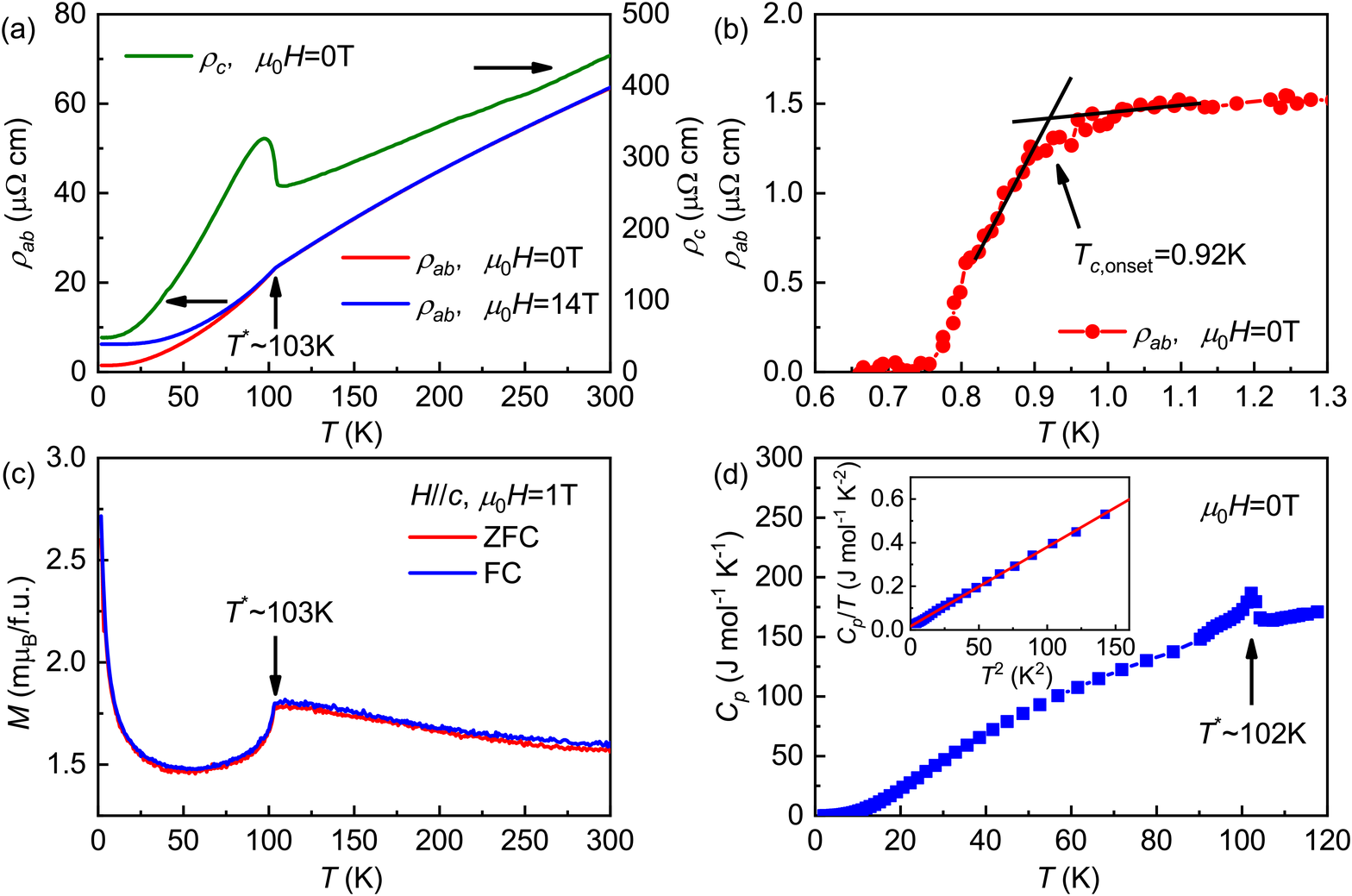}} \vspace*{-0.3cm}
\caption{(a) Temperature dependence of $\rho_{ab}(T)$ and $\rho _{c}(T)$ at zero field and 14 T between 2 K and 300 K. (b) Temperature dependence of zero-field $\rho _{ab}(T)$ below 1.3 K. (c) Temperature dependence of $M(T)$ at $\mu_{0}H=$ 1 T for $H\parallel c$ with ZFC and FC modes. (d) Temperature dependence of $C_{p}(T)$ at zero field between 2 K and 117 K. Inset: $C_{p}/T$ vs. $T^{2}$ at low temperature region. The red solid line represents the linear fit using the formula $C_{p}/T=\gamma+\beta T^{2}$.}
\end{figure}

The charge ordering transition also has a remarkable influence on the magnetic property of RbV$_{3}$Sb$_{5}$. As shown in Fig. 2(c), the magnetization $M(T)$ curve exhibits a relatively weak temperature dependence with a small absolute value above $T^{*}$, reflecting the Pauli paramagnetism of RbV$_{3}$Sb$_{5}$. In contrast, when $T<T^{*}$, there is a sharp drop in the $M(T)$ curve because of the decreased carrier density originating from the partially gapped FS by the charge ordering transition. In addition, the nearly overlapped zero-field-cooling (ZFC) and field-cooling (FC) $M(T)$ curves also suggest that this anomaly should be due to certain density wave transition not an antiferromagnetic one.
Fig. 2(d) shows the temperature dependence of heat capacity $C_{p}(T)$ of RbV$_{3}$Sb$_{5}$ single crystals measured between $T=$ 2 and 117 K at zero field. It can be seen that there is a jump at $\sim$ 102 K, in agreement with the $T^{*}$ obtained from resistivity and magnetization measurements. The jump in $C_{p}(T)$ curve of RbV$_{3}$Sb$_{5}$ is similar to those of KV$_{3}$Sb$_{5}$ and CsV$_{3}$Sb$_{5}$ \cite{Ortiz1,Ortiz2,Ortiz3}, suggesting the same origin of this anomaly of heat capacity from the charge ordering transition. The electronic specific heat coefficient $\gamma$ and phonon specific heat coefficient $\beta$ can be obtained from the linear fit of low-temperature heat capacity using the formula $C_{p}/T=\gamma+\beta T^{2}$ (inset of Fig. 2(d)). The fitted $\gamma$ and $\beta$ is 17(1) mJ mol$^{-1}$ K$^{-2}$ and 3.63(2) mJ mol$^{-1}$ K$^{-4}$, respectively. The latter one gives the Debye temperature $\Theta _{D}=$ 168.9(3) K using the formula $\Theta_{D}=(12\pi ^{4}N\rm{R}/5\beta)^{1/3}$, where $N$ is the number of atoms per formula unit and R is the gas constant. 
The electron-phonon coupling $\lambda _{e-ph}$ can be estimated with the values of $\Theta _{D}$ and $T_{c}$ using McMillan's formula \cite{McMillan},

\begin{equation}
\lambda _{e-ph}=\frac{1.04+\mu ^{\ast }\ln(\Theta _{D}/1.45T_{c})}{(1-0.62\mu ^{\ast })\ln(\Theta _{D}/1.45T_{c})-1.04}
\end{equation}

\noindent where $\mu^{\ast}$ is the repulsive screened Coulomb potential and is usually between 0.1 and 0.15. Assuming $\mu ^{\ast}=$ 0.13, the calculated $\lambda _{e-ph}$ is about 0.489, implying that RbV$_{3}$Sb$_{5}$ is a weakly coupled BCS superconductor \cite{Allen}.

\begin{figure}
\centerline{\includegraphics[scale=0.16]{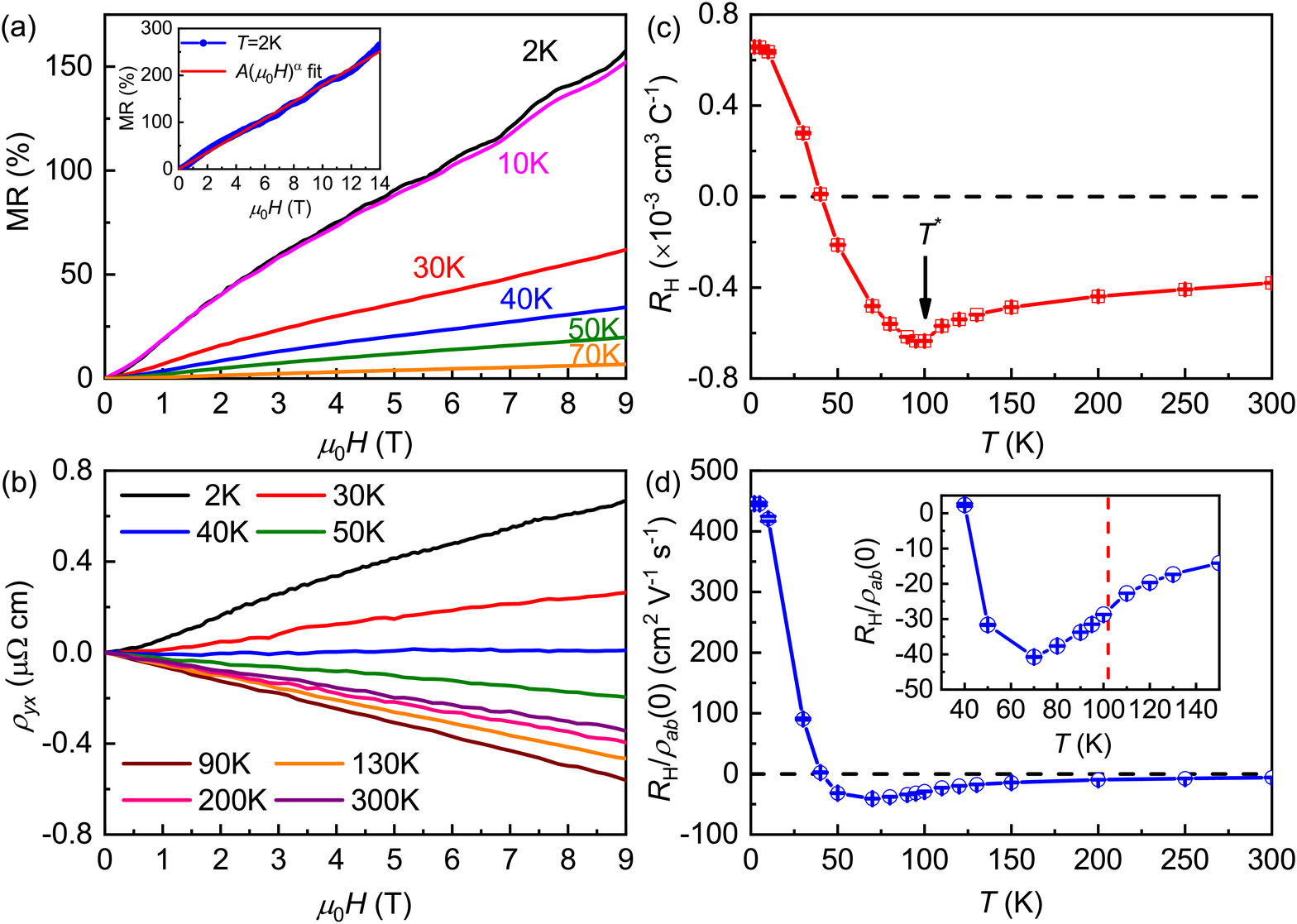}} \vspace*{-0.3cm}
\caption{(a) and (b) Field dependence of MR and $\rho_{yx}(T,\mu_{0}H)$ at various temperatures up to 9 T. Inset of (a) shows the field dependence of MR at 2 K with the field up to 14 T. The red line represents the fit using the formula MR $=A(\mu_{0}H)^{\alpha}$.	
(c) Temperature dependence o $R_{\rm H}(T)$ obtained from the linear fits of $\rho_{yx}(T,\mu_{0}H)$ curves. (d) Temperature dependence of $R_{\rm H}/\rho_{ab}(0)$. Inset: the enlarged part of $R_{\rm H}/\rho_{ab}(0)$ near $T^{*}$ and the vertical red line represents the temperature of $T^{*}$.
}
\end{figure}

The MR [$=(\rho_{ab}(\mu_{0}H)-\rho_{ab}(0))/\rho_{ab}(0)$] of RbV$_{3}$Sb$_{5}$ is negligible above $T^{*}$ and increases gradually below $T^{*}$ (Fig. 3(a)), consistent with the $\rho_{ab}(T)$ data (Fig. 2(a)). At low temperature, the MR does not saturate up to 14 T and the Shubnikov-de Haas (SdH) quantum oscillations (QOs) can be clearly observed at low-temperature and high-field region (inset of Fig. 3(a)). The MR at 2K can be fitted using the formula MR $=A(\mu_{0}H)^{\alpha}$ with $\alpha=$ 1.001(5) (inset of Fig. 3(a)), such linear behavior of MR extends to $T^{*}$, especially at $\mu_{0}H>$ 3 T. 
Fig. 3(b) shows the field dependence of Hall resistivity $\rho_{yx}(T,\mu_{0}H)$ at several typical temperatures. At high temperature, the values of $\rho_{yx}(T,\mu_{0}H)$ are negative with nearly linear dependence on field. When decreasing temperature below 50 K, the $\rho_{yx}(T,\mu_{0}H)$ becomes positive but the linear field dependence is almost unchanged at high-field region. Similar to the MR curves, the SdH QOs appear at low temperatures. The Hall coefficient $R_{\rm H}$ obtained from the linear fits of $\rho_{yx}(T,\mu_{0}H)$ curves are shown in Fig. 3(c). The strong temperature dependence of $R_{\rm H}$ implies that RbV$_{3}$Sb$_{5}$ is a multi-band metal, consistent with theoretical calculations and ARPES measurements of KV$_{3}$Sb$_{5}$ and CsV$_{3}$Sb$_{5}$ \cite{Ortiz1,Ortiz2,YangSY}.
At high temperature, the negative $R_{\rm H}$ suggests that the electron-type carriers are dominant, which could originate from the electron pockets around $\Gamma$ and $K$ points of BZ \cite{Ortiz1,Ortiz2,YangSY}. The most remarkable feature is that the weakly temperature dependent $R_{\rm H}$ starts to decrease rapidly below $T^{*}$ and changes its sign to positive at about 40 K. Such behavior is very similar to the typical CDW materials NbSe$_{2}$ and TaSe$_{2}$ \cite{Naito}, and SDW system GdSi \cite{FengY}.
Both theory and STM results indicate that the $\textbf{q}_{\rm CDW}$ connects the $M$ point when the Fermi level is close to the van Hove filling as in the case of AV$_{3}$Sb$_{5}$ \cite{JiangYX,WangWS,Kiesel,Ortiz1,Ortiz2}. Moreover, there are a band with van Hove singularity and a pair of Dirac-cone like bands near the $M$ point \cite{JiangYX}, which can form hole pockets especially when the $E_{\rm F}$ shifts downward slightly due to the slight Rb deficiency \cite{Ortiz1,Ortiz2}.
Therefore, the charge order may lead to the gap opening of hole bands not electron ones. It seems very peculiar that the $\rho_{ab}(T)$ becomes smaller with positive $R_{\rm H}$ in the charge ordering state even though the portions of hole-type FSs are gapped. Here, we explain this phenomenon in the framework of two-band model. According to the two band model at low-field region \cite{Ziman},

\begin{equation}
R_{\rm H}=\frac{\rho_{yx}}{\mu_{0}H}=\frac{n_{h}\mu_{h}^{2}-n_{e}\mu_{e}^{2}}{e(n_{h}\mu_{h}+n_{e}\mu_{e})^{2}}
\end{equation}

\noindent where $\mu_{e,h}$ and $n_{e,h}$ are the mobilities and densities of electron- and hole-type carriers, respectively. Because of zero-field $\rho_{ab}(0)=1/\sigma_{ab}(0)=1/(n_{h}e\mu_{h}+n_{e}e\mu_{e})$, it has,

\begin{equation}
R_{\rm H}/\rho_{ab}(0)=\frac{n_{h}\mu_{h}^{2}-n_{e}\mu_{e}^{2}}{n_{h}\mu_{h}+n_{e}\mu_{e}}
\end{equation}

The derived $R_{\rm H}/\rho_{ab}(0)$ with the dimension of mobility is shown in Fig. 3(d). According to eq. (3), if the $n_{e}\mu_{e}^{2}$ is much larger than the $n_{h}\mu_{h}^{2}$ which should be the case above $T^{*}$, the $R_{\rm H}/\rho_{ab}(0)$ is negative and the $1/|R_{\rm H}|$ will be close to $n_{e}$, which is about 1.6$\times$10$^{22}$ cm$^{-3}$ at 300 K. 
On the other hand, when the $T$ is just below $T^{*}$, the $\mu_{h}$ may still not increase remarkably and the $n_{h}$ decreases continuously because the FS reconstruction has not finished yet, manifesting from the drops of $M(T)$ curves shown in Fig. 2(c)
This would result in an even negative value of $R_{\rm H}/\rho_{ab}(0)$, which can be clearly seen in the inset of Fig. 3(d).
In contrast, when the $T$ is far below $T^{*}$ ($<$ 70 K), the $n_{h}$ becomes insensitive to temperature and the $\mu_{h}$ may be much larger than the $\mu_{e}$ because both of electron and hole mobilities have the temperature dependence $BT^{-n}$ with different $B$ and $n$ values. It will lead to a sign reversal of $R_{\rm H}/\rho_{ab}(0)$ to positive even the $n_{h}$ is smaller than the value above $T^{*}$. This also explains the even smaller $\rho_{ab}(T)$ below $T^{*}$. 
Since the strongly CDW-like coupled portions of FSs near $M$ point may play a negative role in conductivity above $T^{*}$, the carrier scattering around this area can be effectively reduced when entering charge ordering state, and thus the $\mu_{h}$ can enhance significantly \cite{Valla}.
Similar discussion about the sign change of $R_{\rm H}$ has been developed by Ong for 2D multiband system and applied to Sr$_{2}$RuO$_{4}$ and CDW material 2H-NbSe$_{2}$ \cite{Ong,Mackenzie,LiL}.

\begin{figure}
\centerline{\includegraphics[scale=0.22]{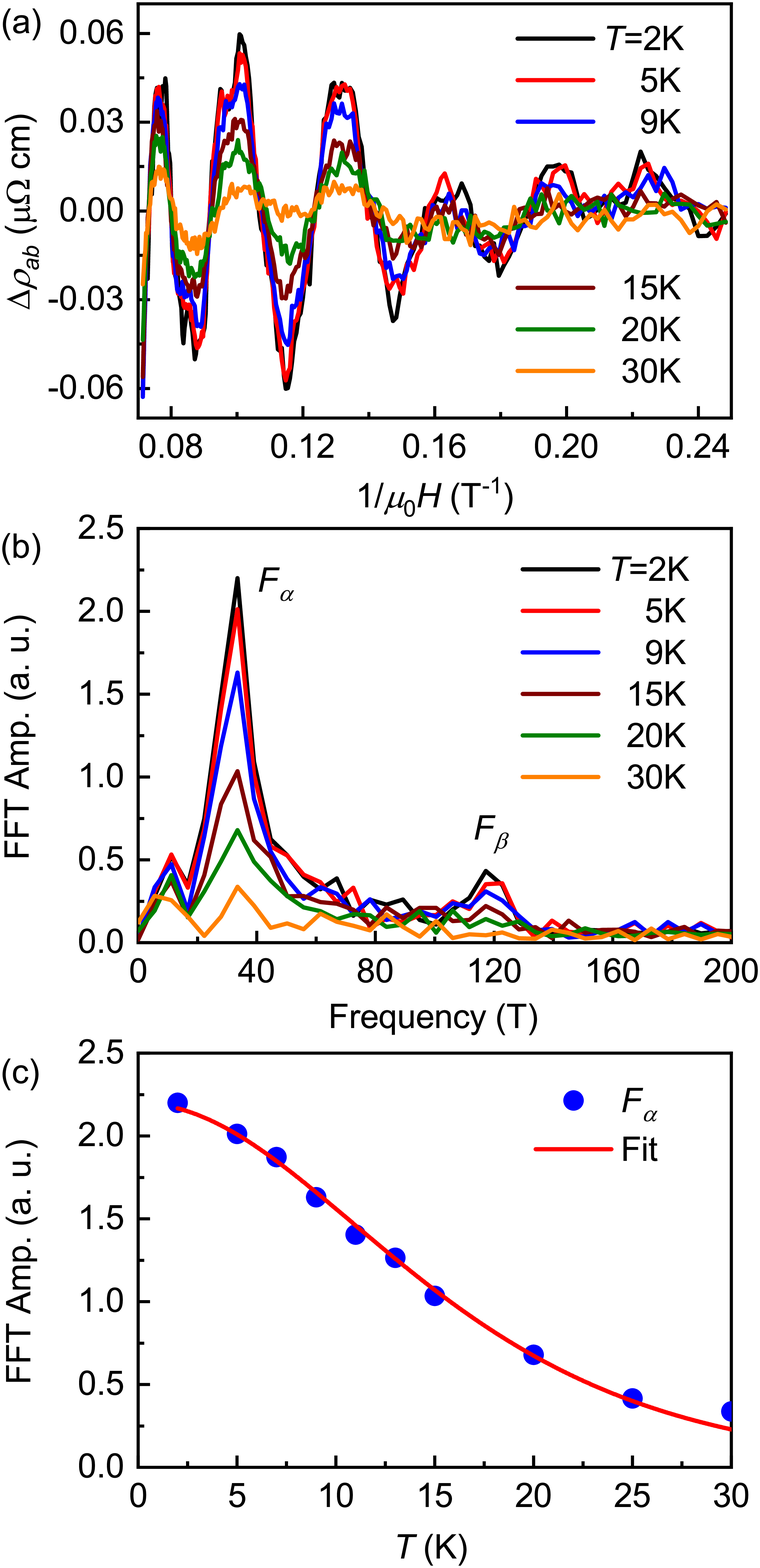}} \vspace*{-0.3cm}
\caption{(a) SdH QOs $\Delta \rho_{ab}=\rho_{ab}-\left\langle\rho_{ab}\right\rangle$ as a function of 1/($\mu_{0}H$) at various temperatures. (b) FFT spectra of the QOs between 4 T and 14 T at various temperatures. (c) The temperature dependence of FFT amplitude of $F_{\alpha}$ frequency. The solid line is the fit using the LK formula to extract the effective mass.}
\end{figure}

Analysis of SdH QOs provides insight on the features of FSs and carriers further. After subtracting the slowly changed part of $\rho_{ab}(\mu_{0}H)$ ($\equiv \left\langle\rho_{ab}\right\rangle$), the oscillation parts of resistivity $\Delta \rho_{ab}=\rho_{ab}-\left\langle\rho_{ab}\right\rangle$ as a function of 1/($\mu_{0}H$) for $H\Vert c$ at several representative temperatures are shown in Fig. 4(a). The amplitudes of QOs decrease with increasing temperature or decreasing field, but still observable at 30 K. The fast Fourier transform (FFT) spectra of the QOs reveal two principal frequencies $F_{\alpha}=$ 33.5 T and $F_{\beta}=$ 117.2 T (Fig. 4(b)). Both of them are slightly smaller than those in KV$_{3}$Sb$_{5}$ \cite{YangSY}, indicating RbV$_{3}$Sb$_{5}$ has smaller extremal orbits of FSs than KV$_{3}$Sb$_{5}$. According to the Onsager relation $F=(\hbar/2\pi e)A_{F}$ where $A_{F}$ is the area of extremal orbit of FS, the determined $A_{F}$ is 0.0032 and 0.011 \AA$^{-2}$ for $\alpha$ and $\beta$ extremal orbits, respectively. These $A_{F}$s are very small, taking only about 0.0934 \% and 0.321 \% of the whole area of BZ in the $k_{x}-k_{y}$ plane when taking the lattice parameter $a=$ 5.4715 \AA\ \cite{Ortiz1}.
The effective mass $m^{*}$ can be extracted from the temperature dependence of the amplitude of FFT peak using the Lifshitz-Kosevich (LK) formula,

\begin{equation}
\Delta\rho_{ab} \propto \frac{X}{\sinh X}
\end{equation}

\noindent where $X=2\pi^{2}k_{B}T/\hbar\omega_{c}=14.69m^{*}/\mu_{0}H_{\rm avg}$ with $\hbar\omega_{c}$ being the cyclotron frequency and $\mu_{0}H_{\rm avg}$ (= 9 T) being the average value of the field window used for the FFT of QOs \cite{Shoenberg,Rhodes}. As shown in Fig. 4(c), the temperature dependence of FFT amplitude of $F_{\alpha}$ can be fitted very well using eq. (4) and the obtained $m^{*}$ is 0.091(2) $m_{e}$, where $m_{e}$ is the bare electron mass. This value is even smaller than that in KV$_{3}$Sb$_{5}$ (0.125 $m_{e}$ for the $\alpha$ orbit) \cite{YangSY}. The small extremal cross sections of FSs accompanying with such light $m^{*}$ could be related to the highly dispersive bands near either $M$ point or along the $\Gamma-K$ path of BZ \cite{Ortiz1,Ortiz2,YangSY,JiangYX}.

In summary, we carried out the detailed study on physical properties of RbV$_{3}$Sb$_{5}$ single crystals grown by the self-flux method. RbV$_{3}$Sb$_{5}$ single crystals exhibit a superconducting transition at $T_{c,\rm onset}$ = 0.92 K with a weakly coupling strength, accompanying with anomalies of properties at $T^{*}\sim$ 102 - 103 K. The high-temperature anomaly could be related to the formation of charge ordering state and it results in the sign change of $R_{\rm H}$, which can be partially ascribed to the enhancement of mobility for hole-type carriers due to the reduced carrier scattering by the gapping of strongly CDW-like coupled portions of FSs. Furthermore, there are some very small FSs with rather low $m^{*}$, indicating the existence of highly dispersive bands near $E_{\rm F}$ in RbV$_{3}$Sb$_{5}$. Moreover, due to the similar electronic structure of RbV$_{3}$Sb$_{5}$ to KV$_{3}$Sb$_{5}$ and CsV$_{3}$Sb$_{5}$ \cite{Ortiz1}, RbV$_{3}$Sb$_{5}$ should also be a candidate of $Z_{2}$ topological metal. Therefore, the V-based kagome metals AV$_{3}$Sb$_{5}$ provide a unique platform to explore the interplay between nontrivial band topology, electronic correlation and possible unconventional superconductivity. 

This work was supported by  National Natural Science Foundation of China (Grant No. 11822412 and 11774423), Ministry of Science and Technology of China (Grant No. 2018YFE0202600 and 2016YFA0300504), Beijing Natural Science Foundation (Grant No. Z200005), and Fundamental Research Funds for the Central Universities and Research Funds of Renmin University of China (RUC) (Grant No. 18XNLG14 and 19XNLG17).






$^{\dag}$ Q.W.Y, Z.J.T. and C.S.G. contributed equally to this work.


\end{document}